\documentclass[twoside]{article}
\usepackage{fleqn,espcrc2}

\usepackage{graphicx}
\usepackage[figuresright]{rotating}

\def\simlt{\stackrel{<}{{}_\sim}}
\def\simgt{\stackrel{>}{{}_\sim}}

\def\mpla#1#2#3{{\it Mod.~Phys.~Lett.} {\bf#1} (#2) #3}

\newcommand{\npb}[3]{\emph{ Nucl.~Phys.} \textbf{B#1} (#2) #3}   
\newcommand{\plb}[3]{\emph{ Phys.~Lett.} \textbf{B#1} (#2) #3}   
\newcommand{\prd}[3]{\emph{ Phys.~Rev.} \textbf{D#1} (#2) #3}   
\newcommand{\prl}[3]{\emph{ Phys.~Rev.~Lett.} \textbf{#1} (#2) #3}

\newcommand{\pr}[3]{\emph{ Phys.~Rep.} \textbf{#1} (#2) #3}   
\newcommand{\rmp}[3]{\emph{ Rev.~Mod.~Phys.} \textbf{#1} (#2) #3}   
\newcommand{\hpa}[3]{\emph{ Helv.~Phys.~Acta} \textbf{#1} (#2) #3}   
\newcommand{\ap}[3]{\emph{ Ann.~Phys.} \textbf{#1} (#2) #3} 
  
\newcommand{\jhep}[3]{\emph{ JHEP} \textbf{#1} (#2) #3}  
\newcommand{\Tr}{\mathop{\rm Tr}}

\newcommand{\hS}{\widetilde{\mathcal{H}}}

\newcommand{\AmS}{{\protect\the\textfont2
  A\kern-.1667em\lower.5ex\hbox{M}\kern-.125emS}}

\title
{
\begin{flushright}   
\normalsize{     
IEM-FT-208/00\\ 
hep-ph-/0101230\\ 
} 
\end{flushright}
Electroweak Baryogenesis and the Higgs and Stop 
masses~\thanks{Based on talks given at: {\it Strong Electroweak Matter},
Centre de Physique Th\'eorique, Universit\'e de Marseille, Marseilles 
(France), June 14-17, 2000; and, {\it Thirty years of supersymmetry},
Theoretical Physics Institute, School of Physics and Astronomy, Minneapolis,
Minnesota (USA), October 16-27, 2000.}}

\author{Mariano Quir\'os\address[MCSD]{Instituto de Estructura de la Materia, 
Serrano 123 E-28006, Spain}}

\begin{document}

\begin{abstract}
In this talk we review the actual situation concerning
electroweak phase transition and baryogenesis in the minimal supersymmetric
extension of the Standard Model. A strong enough phase transition requires 
light Higgs and stop eigenstates. For a Higgs mass in the range 110--115 GeV,
there is a stop window in the range 105--165 GeV. If the Higgs is heavier
than 115 GeV, stronger constrains are imposed on the space of supersymmetric
parameters. A baryon-to-entropy ratio is generated by the chargino 
sector provided that the $\mu$ parameter has a CP-violating phase larger
than $\sim$ 0.04.
\vspace{1pc}
\end{abstract}

\maketitle

\section{Introductory remarks}

Electroweak baryogenesis~\cite{baryogenesis} is an appealing mechanism to 
explain the observed, Big Bang Nucleosynthesis (BBN), 
value of the baryon-to-entropy ratio~\cite{BBN}, 
$\eta_{\rm BBN}\equiv n_B/s\sim 4\times 10^{-10}$, at the
electroweak phase transition~\cite{reviews}, that can be tested at 
present and future 
high-energy colliders. Although the Standard Model (SM) contains all the
necessary ingredients~\cite{baryogenesis} for a successful baryogenesis, 
it fails in providing enough baryon asymmetry. In particular it has been
proven by perturbative~\cite{pert} and
non-perturbative~\cite{nonpert} methods that, for Higgs masses allowed by
present LEP bounds~\cite{Higgs}, 
the phase transition is too weakly first order, or does
not exist at all, and any
previously generated baryon asymmetry would be washed out after the phase
transition. On the other hand the amount of CP violation arising from the CKM
phase is too small for generating the observed baryon asymmetry~\cite{CPSM}.
Therefore electroweak baryogenesis requires physics beyond the Standard Model
at the weak scale.

Among the possible extensions of the Standard Model at the weak scale, its
minimal supersymmetric extension (MSSM) is the best motivated one. It
provides a technical solution to the hierarchy problem and has deep roots 
in more fundamental theories unifying gravity with the rest of interactions.
As for
the strength of the phase transition~\cite{mariano}, 
a region
in the space of supersymmetric parameters has been 
found~\cite{1mssm}-\cite{17mssm} where the phase transition is strong 
enough to let
sphaleron interactions go out of equilibrium after the phase transition and
not erase the generated baryon asymmetry. This region (the so-called 
light stop scenario) provides values of the
lightest Higgs and stop eigenstate masses which are being covered at LEP and
Tevatron colliders.

The MSSM has new violating phases that can drive enough
amount of baryon asymmetry.
Several computations have been performed~\cite{hn}-\cite{plus} 
in recent years, showing
that if the CP-violating phases associated with the chargino
mass parameters are not too small, these sources may lead to 
acceptable values of the baryon asymmetry. In this talk I will present some
aspects of a recent computation~\cite{recent} of the CP-violating 
sources in the chargino sector which improves  the computation of 
Ref.~\cite{CQRVW} in two main aspects. On the one hand, 
instead of computing the temporal
component of the current in the lowest order of Higgs 
background insertions, we compute all current components by 
performing a resummation of the 
Higgs background insertion contributions to all order in perturbation
theory. The resummation is essential since it leads to a
proper regularization of the resonant contribution to the
temporal component of the current found in Ref.~\cite{CQRVW}
and leads to contributions which are not suppressed for
large values of the charged Higgs mass. On the other hand we consider, in
the diffusion equations, the contribution of Higgsino number violating
interaction rate~\cite{plus} from the Higgsino $\mu$ term in the lagrangian, 
$\Gamma_\mu$, that was considered in our previous calculations in the limit
$\Gamma_\mu/T\to\infty$. 

\section{The phase transition}

The possibility of achieving, in the MSSM, a strong-enough phase transition 
for not washing out any previously generated baryon asymmetry, characterized
by the condition
\begin{equation}
v(T_c)/T_c\simgt 1\quad ,
\label{condicion}
\end{equation}
has been recently strengthened by three facts:
\begin{itemize}
\item
The presence of light $\widetilde{t}_R$ (with small mixing $\widetilde{A}_t$)
considerably enhances the strength of the phase
transition~\cite{mariano,1mssm}. This is the so-called light
stop scenario. 
\item
Two-loop corrections enhance the phase transition in the SM~\cite{pert},
and in the MSSM~\cite{7mssm,8mssm,9mssm}.
\item
The validity of perturbation theory, and in particular the results of two-loop
calculations, for the light stop scenario has been recently confirmed by
non-perturbative results~\cite{15mssm,16mssm}. In fact, non-perturbative 
results lead to a stronger phase transition than perturbative ones, by
$\sim 10-15$\%.
\end{itemize}

However the price the light stop scenario has to pay is that it may require
moderately negative values of the supersymmetric parameter
$m_U^2\equiv-\widetilde{m}_U^2$ and then, apart from the electroweak minimum
along the Higgs ($\phi$) direction, another color breaking minimum along the
$U\equiv\widetilde{t}_R$ direction might appear. Therefore both
directions should be studied at finite temperature.

To systematically analyze the different possibilities we have
computed the two-loop effective potential along the $\phi$ and
$U$ directions and compared their cosmological evolutions with
$T$. The two-loop effective potential along the $\phi$ direction
was carefully studied in Ref.~\cite{7mssm}. The one-loop
correction is dominated by the exchange of the top/stop sector
while the two-loop effective potential is given by two-loop 
diagrams with stops and gluons, as well as one-loop diagrams with
the stop thermal counterterm. The two-loop effective potential
along the $U$ direction was studied in Refs.~\cite{9mssm}. 
The mass spectrum is given in Table 1, where we have 
considered small mixing $\widetilde{A}_t/m_Q$ and large gluino masses.
\begin{table}[t]
\caption{Mass spectrum along the $U$ direction.
\label{spectrum}}
\vspace{0.2cm}
\begin{center}
\footnotesize
\begin{tabular}{|c|c|c|}
\hline &&\\
field & d.o.f. &  mass$^2$ \\ \hline &&\\
4 gluons &12& $g_s^2\,U^2/2$\\
1 gluon &3& $2g_s^2\, U^2/3$ \\
1 $B$ gauge boson &3& $\, g'^2 U^2/9$\\ 
5 squark-goldstones &5& $m_U^2+g_s^2\, U^2/3$ \\
1 squark &1 & $m_U^2+g_s^2\, U^2$ \\
\hline &&\\
4 $\widetilde{Q}_L$-Higgs &4& $-m_H^2/2+h_t^2\sin^2\beta U^2$\\
2 Dirac fermions ($t_L,\widetilde{H}$) &8&$\mu^2+h_t^2 U^2$\\
\hline
\end{tabular}
\end{center}
\end{table}
The one-loop diagrams correspond to the propagation of gluons,
squarks and Higgses as well as Dirac fermions. The leading
two-loop contributions correspond to sunset and figure-eight
diagrams with the fields of Table 1 propagating, as well as
one-loop diagrams with thermal counterterm insertions
corresponding to gluons and squarks.

For a given value of the supersymmetric parameters we have
computed $T_c$, the critical temperature along the $\phi$
direction, and $T_c^U$, the critical temperature along the $U$
direction. Therefore, four different situations can arise:
\begin{itemize}
\item {\bf a)}
$T_c^U<T_c$ and $\widetilde{m}_U<m_U^{c}\equiv 
\left(m_H^2 v^2 g_s^2/12\right)^{1/4}$~\cite{1mssm}. In this case the phase transition proceeds first along the $\phi$ direction and the field remains at
the electroweak minimum forever. This region is called {\em stability} region.
\item {\bf b)}
$T_c^U<T_c$ and $\widetilde{m}_U>m_U^c$. In this region the electroweak 
minimum is metastable ({\em metastability} region). It can be physically 
acceptable provided that its lifetime is larger than the age of the universe
at this temperature: $\Gamma_{\phi\rightarrow U}<H$.
\item {\bf c)}
$T_c^U>T_c$ and $\widetilde{m}_U<m_U^c$. In this case the $U$-phase transition
happens first and therefore it would be 
physically acceptable provided that the
lifetime is shorter than the age of the universe: 
$\Gamma_{U\rightarrow \phi}>H$. This region is called {\em two
step} region and has been excluded~\cite{17mssm}. 
\item {\bf d)}
$T_c^U>T_c$ and $\widetilde{m}_U>m_U^c$. This is the region of 
{\em instability} of the electroweak minimum. It is absolutely excluded.
\end{itemize}

%
\begin{figure}[htb]
\vspace{9pt}
\includegraphics*[scale=0.32]{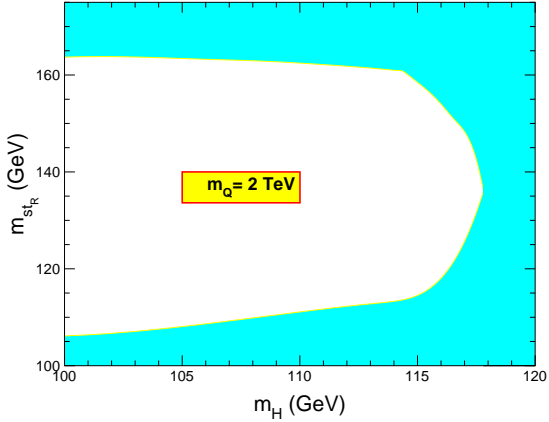}
\caption{The absolute region of stability in the ($m_H,m_{\widetilde{t}_R}$) plane, for $m_Q=2$ TeV. The allowed region is inside the solid lines.}
\label{fig:bound}
\end{figure}
Fig.~\ref{fig:bound} shows the absolute region of stability
for $m_Q\leq 2$ TeV, where we can see an absolute
upper bound on the Higgs mass $\sim$117 GeV. In view of the non-perturbative
results on the phase transition~\cite{nonpert},
we have used in Fig.~\ref{fig:bound} the
condition $v(T_c)\simgt 0.9 T_c$ in our perturbative two-loop calculation.

\section{The CP-violating chargino currents}

Our aim in this section is to compute the 
Green functions for the charged gaugino-Higgsino system
$\widetilde{h}_c$-$\widetilde{W}_c$,
describing the propagation of these fermions in the presence
of a bubble wall. The bubble wall is assumed to be located at the
space-time point $z$, where there is a non-trivial 
background of the MSSM Higgs fields, $H_i(z)$, which carries
dimensionful CP-violating couplings to charginos.
We shall use these Green functions to
compute the left-handed and right-handed currents corresponding to
\begin{equation}
\psi_R(x)=\left(
\begin{array}{c}
\widetilde{W}^+\\ \widetilde{h}_2^+
\end{array}\right),
\qquad
\psi_L(x)=\left(
\begin{array}{c}
\widetilde{W}^-\\ \widetilde{h}_1^-
\end{array}\right) \ .
\end{equation}
at the point $z$.

Since the mass matrix depends on the 
space-time coordinates, and we must identify the free and perturbative parts
out of it,
in order to make such a selection we will expand the mass matrix around the
point $z^\mu\equiv (\vec{r},t)$ (the point where we are calculating the 
currents in the plasma frame) up to first order 
in derivatives as,
\begin{equation}
\label{Mexp}
M(x)=M(z)+(x-z)^\mu M_\mu(z) \ ,
\end{equation}
where we use the notation $M_\mu(z)\equiv\partial M(z)/\partial z^\mu$.

The chargino mass matrix in (\ref{Mexp}) is given by
\begin{equation}
\label{masach}
M(z)=\left(
\begin{array}{cc}
M_2 & u_2(z) \\
u_1(z) & \mu_c
\end{array}
\right)
\end{equation}
where we have defined $u_i(z)\equiv g H_i(z)$. 
The mass eigenvalues are given by
\begin{eqnarray}
\label{eigenval}
m_1(z)= \frac{\left(\Delta+\Lambda+u_1^2(z)\right)M_2+u_1(z) u_2(z)\mu^*_c}
{\sqrt{(\Delta+\Lambda)(\bar\Delta+\Lambda)}}
\nonumber\\
m_2(z)= \frac{\left(\Delta+\Lambda-u_2^2(z)\right)\mu_c-u_1(z) u_2(z)M_2}
{\sqrt{(\Delta+\Lambda)(\bar\Delta+\Lambda)}} \nonumber\ .
\end{eqnarray}
where field redefinitions have been made in order to make
the Higgs vacuum expectation values, as well as the weak gaugino
mass $M_2$, real, and
\begin{eqnarray}
\label{defin}
\Delta&=&(M_2^2-|\mu_c|^2-u_1^2+u_2^2)/2 \nonumber\\
\bar\Delta&=&(M_2^2-|\mu_c|^2-u_2^2+u_1^2)/2 \nonumber\\
\Lambda&=&\left(\Delta^2+\left|M_2\,u_1+\mu^*_c\, u_2 \right|^2\right)^{1/2}
\end{eqnarray}

The vector and axial Higgsino currents can now be defined as:
\begin{eqnarray}
\label{chcurr}
j_{H,h}^\mu(z)&=&\lim_{x,y\to z}\left\{\Tr\left[P_2\sigma^\mu 
S_\psi^{RR}(x,y;z)\right]\right.\nonumber\\
&\pm& \left.
\Tr\left[P_2\overline{\sigma}^\mu S_\psi^{LL}(x,y;z)\right]\right\}
\end{eqnarray}
where $P_2=(\sigma_0-\sigma_3)/2$ is a projection operator and $S_\psi(x,y;z)$
are the Green functions in the weak eigenstate basis after making the 
expansion (\ref{Mexp}) and a resummation to all orders in $M(z)$.

The detailed calculation of the currents
can be found in Ref.~\cite{recent}. In particular 
only the time component is relevant for the diffusion equations
of next section where we will use as sources of Higgs densities
$\gamma_{H,h}\simeq
\Gamma_{\widetilde{\cal H}}\, j_{H,h}^0(z)$, and 
$\Gamma_{\widetilde{\cal H}}\sim \alpha_W\, T$
is the inverse typical thermalization time. In this way we obtain for the
vector and axial sources the expressions:

\begin{eqnarray}
\label{sourceH}
&\gamma_{H}\simeq -\, 2\, v_\omega\, g^2\, 
\Gamma_{\hS} \, {\rm Im}(M_2\mu_c) &
\\
&\left\{ H^2(z)\, \beta'(z)\,
\left[ \mathcal{F}(z)+\mathcal{G}(z)
\right]\right.&
\nonumber\\
&+ g^2\, H^2(z)\cos 2\beta(z)\left[
H(z) H'(z)\sin 2\beta(z)\right. &
\nonumber\\
&+\left.\left.H^2(z)\cos 2\beta(z) \beta'(z)
\right] \mathcal{H}(z) \right\}\nonumber &
\end{eqnarray}
\begin{eqnarray}
\label{sourceh}
&\gamma_{h}\simeq \ 2\, v_\omega\, g^2\, 
\Gamma_{\hS} \, {\rm Im}(M_2\mu_c)&\nonumber\\
&\left[
H(z) H'(z)\sin 2\beta(z)+H^2(z)\cos 2\beta(z) \beta'(z)
\right]\nonumber &\\
&\left\{ \mathcal{K}(z)
+2\left(\Delta+\bar\Delta\right)\mathcal{H}(z)\right\}&\nonumber \ .
\end{eqnarray}
where the functions $\mathcal{F,G,H,K}$ are defined as:
\begin{eqnarray}
\label{integralF}
&{\cal F}(z)=\frac{1}{6\pi^2}{\rm Re}\int_0^\infty dp^0\,(1+2\, f)&\nonumber\\
&{\displaystyle \left(\frac{1}{z_1+z_2}\right)^3 }&
\end{eqnarray}
\begin{eqnarray}
\label{integralG}
&&{\cal G}(z)=\frac{1}{3\pi^2}{\rm Re}\int_0^\infty\,dp^0\,  p^0 f'
\nonumber\\
&&\left\{\left(\frac{1}{z_1+z_2}\right)^3
-\frac{3}{|m_1(z)|^2-|m_2(z)|^2}\right.\nonumber\\
&& \left[\frac{z_1}{|m_1(z)|^2-|m_2(z)|^2-4 i 
\Gamma_{\widetilde H}\ p^0}\right.\nonumber\\
&&\left.\left.+
\frac{z_2}{|m_1(z)|^2-|m_2(z)|^2+4 i \Gamma_{\widetilde H}\ p^0}\right]\right\}
\end{eqnarray}
\begin{eqnarray}
\label{integralH}
&{\displaystyle
\mathcal{H}(z)=\frac{1}{8\pi^2}{\rm Re}\int_0^\infty dp^0\,(1+2\, f)}&
\nonumber\\
&{\displaystyle \frac{1}{z_1\, z_2}\left(\frac{1}{z_1+z_2}\right)^3} &
\end{eqnarray}
\begin{eqnarray}
\label{integralK}
&{\displaystyle
\mathcal{K}(z)=-\,\frac{1}{4\,\pi^2}{\rm Re}\int_0^\infty dp^0\,(1+2\, f)}&
\nonumber\\
&{\displaystyle \frac{1}{z_1\,z_2}\left(\frac{1}{z_1+z_2}\right)}&\ .
\end{eqnarray}
$f\equiv -n_F(|p^0|)$, where $n_F$ is the Fermi-Dirac distribution function
and $z_i$ is defined as 
\begin{equation}
\label{zetas}
z_i(p^0)=\sqrt{p^0\left(p^0+2\, i \Gamma_{\widetilde H}\right)-|m_i(z)|^2}
\end{equation}
\begin{figure}[htb]
\vspace{9pt}
\includegraphics*[scale=0.28]{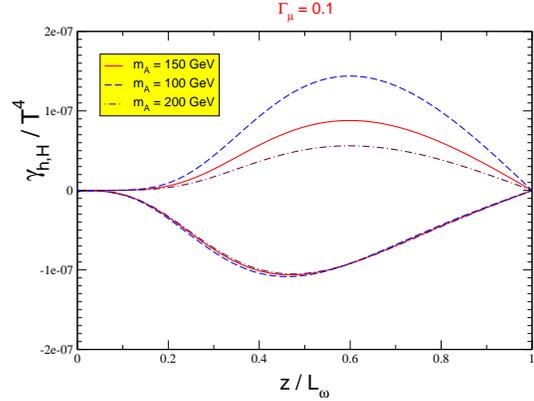}
\caption{Plot of the sources $\gamma_{H,h}$, for the values of supersymmetric
parameters specified in the text, as functions of $z/L_\omega$.}
\label{fig:sources}
\end{figure}
with positive real and imaginary parts satisfying ${\rm Re}(z_i)
=\Gamma_{\widetilde H}\ p^0/{\rm Im}(z_i)$. 
Notice that the sources are proportional to the wall
velocity $v_\omega$, and so die when the latter goes to zero, which is
a physical requirement. 

In Fig.~\ref{fig:sources} we plot the sources $\gamma_{H,h}$ for a chosen set
of supersymmetric and bubble parameters. In particular the wall velocity 
is chosen as $v_\omega=0.05$ and the bubble wall width as $L_\omega=20/T$ 
which is suggested by detailed numerical analyses of bubble 
formation~\cite{11mssm,velocity}. For the supersymmetric parameters we choose
$m_Q=1.5$ TeV, $A_t=0.5$ TeV, $M_2=\mu=200$ GeV, 
$\tan\beta= 20$ and three different values
of $m_A=100,\, 150,\, 200$ GeV, corresponding to the dashed, solid and 
dot-dashed curves of Fig.~\ref{fig:sources}. From it we can see two main 
features. On the one hand, $\gamma_H$ is very sensitive to the value of
$m_A$, and so to the corresponding value of $\Delta\beta$, as expected, and
decreases when $m_A$ increases. On the other hand, $\gamma_h$ is dominant
only for large values of $m_A$, such that the $\Delta\beta$ suppression of
$\gamma_H$ is stronger. But it is never overwhelming the contribution of 
$\gamma_H$, because it is $\tan\beta$ suppressed.

We also computed the bubble solutions of the MSSM~\cite{11mssm} using 
the two-loop effective potential and we have 
confirmed the goodness of the thick wall approximation ($L_\omega T_c
\gg 10\, v_\omega$).

%
\begin{figure}[htb]
\includegraphics*[scale=0.4]{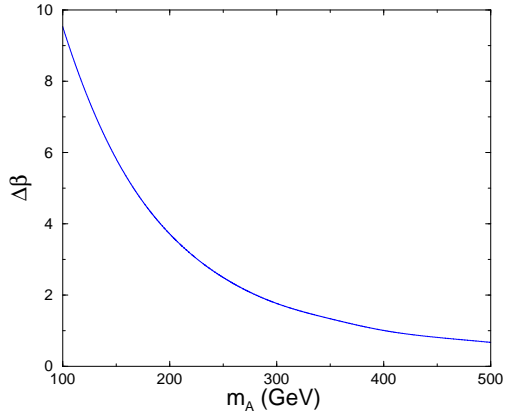}
\caption{$\Delta\beta$, in units of $10^{-3}$, 
as a function of $m_A$ for the values of the
supersymmetric and bubble parameters indicated in the text.}
\label{fig:beta}
\end{figure}
%

In Fig.~\ref{fig:beta}, we plot $\Delta\beta(T)$ for the following values of
the supersymmetric: $M_2=\mu=200$ GeV, $A_t=500$ GeV, $m_Q=1.5$ TeV. 
Notice that the value of $T_c$, as well as $\Delta\beta$, is different from
point to point, for different values of $m_A$. We see from Fig.~\ref{fig:beta}
that $\Delta\beta$ goes from $\sim 10^{-2}$ for $m_A=100$ GeV, to 
$\sim 10^{-3}$ for $m_A=500$ GeV. This fact will have an important impact 
on the mechanism of baryon asymmetry.

\section{The diffusion equations}

To evaluate the baryon asymmetry generated in the broken phase we need
to first compute the density of left-handed quarks and leptons, $n_L$,
in front of the
bubble wall (in the symmetric phase). These chiral densities are
the ones that induce
weak sphalerons to produce a net
baryon number. Since, in the present scenario, there is essentially no lepton
asymmetry, the density to be computed in the symmetric 
phase is $n_L=n_{Q}+\sum_{i=1}^2 n_{Q_i}$
where the density of a chiral supermultiplet 
$Q \equiv (q,\tilde q)$ 
is understood as the sum of densities of particle
components, assuming the supergauge interactions to be in thermal 
equilibrium, $n_Q=n_q+n_{\tilde q}$. If the system is near thermal equilibrium,
particle densities, $n_i$, are related to the local chemical potential,
$\mu_i$ by the relation $n_i=k_i \mu_i T^2/6$, where $k_i$ are statistical 
factors equal to 2 (1) for bosons (fermions). In fact, assuming 
that all quarks have nearly
the same diffusion constant it turns out that~\cite{hn},
$n_L=5\, n_Q+\, 4\, n_T$.

One can write now a set of diffusion equations involving $n_Q$, $n_T$, 
$n_{H_1}$ (the density of $H_1\equiv(h_1,\tilde h_1)$) and $n_{H_2}$ 
(the density of $\bar H_2\equiv(\bar h_2,\tilde{\bar h}_2)$), and
the particle number changing rates and CP-violating source terms
discussed above.
In the bubble wall frame, and ignoring the curvature of the bubble wall,
all quantities become functions of $z\equiv r+ v_\omega t$. In the limit of
fast Yukawa coupling $\Gamma_Y$ and strong sphaleron $\Gamma_{ss}$
rates, we can write the diffusion equations as:
\begin{eqnarray}
&&v_\omega\left[n'_Q+2\, n'_T-n'_H\right]= D_q\left[n''_Q+2\,n''_T\right]
\nonumber\\
&&-D_h\, n''_H
+\Gamma_m\left[\frac{n_Q}{k_Q}-\frac{n_T}{k_T} \right]
\Gamma_h
\frac{n_H}{k_H}-\gamma_{H}\nonumber
\\
\label{nh2}
&&v_\omega\left[n'_Q+2\, n'_T-n'_h\right]= D_q\left[n''_Q+2\,n''_T\right]
\nonumber\\
&&-D_h\, n''_h
+\Gamma_m\left[\frac{n_Q}{k_Q}-\frac{n_T}{k_T} \right]\nonumber\\
&&+
\left[\Gamma_h+4\, \Gamma_\mu\right] 
\frac{n_h}{k_H}-\gamma_{h}
\ .
\end{eqnarray}
where $n_Q$ and $n_T$ are replaced by the (approximate) explicit solutions 
\begin{eqnarray}
\label{QT}
n_Q&=&\ \frac{k_Q\left(9 k_T-k_B\right)}{k_H\left(k_B+9 k_Q
+9 k_T\right)}\
(n_H+\, n_h)\nonumber\\ 
n_T&=&-\ \frac{k_T\left(9 k_Q+2 k_B\right)}{k_H\left(k_B+9 k_Q
+9 k_T\right)}\
(n_H+\,  n_h)\ .
\end{eqnarray}
$D_q\sim 6/T$ and $D_h\sim 110/T$ are the corresponding diffusion constants
in the quark and Higgs sectors,
$n_H\equiv n_{H_2}+n_{H_1}$, $n_h\equiv n_{H_2}-n_{H_1}$,
$k_H\equiv k_{H_1}+k_{H_2}$, and $\Gamma_\mu$ corresponds to the 
$\mu_c \tilde{H}_1 \tilde{H_2}$ term in the Lagrangian.
There are also the Higgs number
violating and axial top number violation processes, induced by the 
Higgs self interactions and by top quark mass effects,
with rates $\Gamma_h$
and $\Gamma_m$, respectively, that are only active in the broken phase.
\begin{figure}[htb]
\vspace{9pt}
\includegraphics*[scale=0.28]{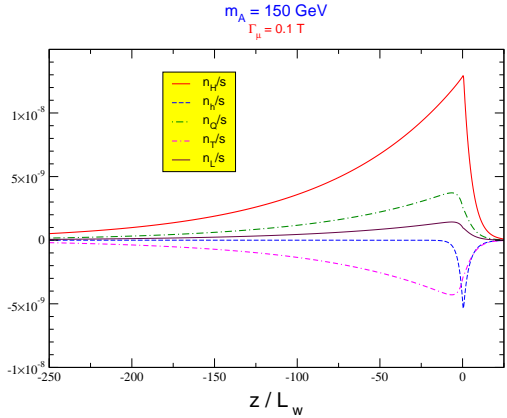}
\caption{Plot of the different density-to-entropy ratios for the values of
supersymmetric parameters specified in the text.}
\label{fig:densities}
\end{figure}

\begin{figure}[htb]
\vspace{9pt}
\includegraphics*[scale=0.28]{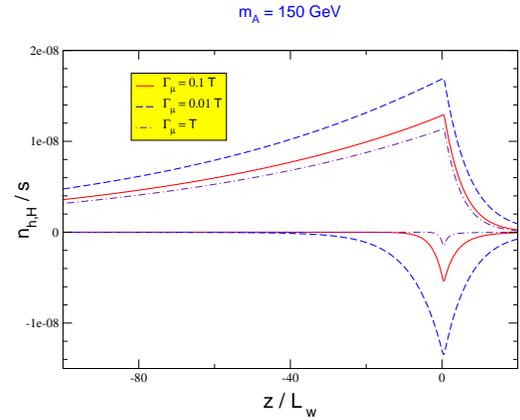}
\caption{Plots of densities-to-entropy ratios $n_{H,h}/s$
as functions of $z/L_\omega$ for different values of $\Gamma_\mu$. }
\label{fig:gammamu}
\end{figure}

The system of equations (\ref{nh2}) has been solved numerically in
Ref.~\cite{recent}, where also a good enough analytical approximation is
provided. In Fig.~\ref{fig:densities} we plot the numerical solution 
corresponding to the same set of supersymmetric and bubble parameters as in
Fig.~\ref{fig:sources}. All densities diffuse along the
symmetric phase ($z<0$), where weak sphalerons are active, which is essential
for the left-handed quark asymmetry to bias the weak sphalerons
to violate baryon number.
In particular, we can see from Fig.~\ref{fig:densities}
that the density $n_H$ is larger than $n_h$, for $\Gamma_\mu=0.1\, T$. In 
previous analyses~\cite{hn,CQRVW} the limit $\Gamma_\mu\to\infty$ was
implicitly assumed and
$n_h$ was completely neglected. 
Fig.~\ref{fig:densities} shows that this is not such a bad approximation.
Furthermore, the relative importance of $n_H$ and $n_h$ is shown for 
different values of $\Gamma_\mu=0.01\,T,\, 0.1\, T,\,  T$
in Fig.~\ref{fig:gammamu}. We can see that
while for  $\Gamma_\mu=0.01\, T$ $n_h$ is sizeable, for $\Gamma_\mu=\, T$
it is negligible in good agreement with our previous results~\cite{CQRVW}.

\section{The baryon asymmetry}

In this section we present the numerical results of the baryon asymmetry
computed in the previous section and, in particular, of the baryon-to-entropy
ratio $\eta\equiv n_B/s$, where the entropy density is given by
$s=\frac{2\pi^2}{45}g_{eff}\, T^3$, with
$g_{eff}$ being the effective number of relativistic degrees of freedom.

Since we assume the sphalerons are inactive inside the bubbles, the
baryon density is constant in the broken phase and satisfies,  
in the symmetric phase, an equation where $n_L$ acts as a 
source~\cite{hn} and there is an explicit sphaleron-induced relaxation 
term~\cite{recent}
\begin{equation}
\label{ecbaryon}
v_\omega n'_B(z)=-\theta(-z)\left[n_F \Gamma_{ws} n_L(z)
+ \mathcal{R}n_B(z)\right]
\end{equation}
where $n_F=3$ is the number of families and $\mathcal{R}
=\,\frac{5}{4}\, n_F\, \Gamma_{ws}$ is the relaxation
coefficient.
Eq.~(\ref{ecbaryon}) can be solved analytically and gives, in the broken
phase $z\ge 0$, a constant baryon asymmetry,
\begin{equation}
\label{nBsol}
n_B=-\,\frac{n_F \Gamma_{ws}}{v_\omega} \int_{-\infty}^0
dz\, n_L(z)\ e^{z\mathcal{R}/v_\omega}\ .
\end{equation}

The profiles $H(z)$, $\beta(z)$ have been accurately computed in the 
literature~\cite{11mssm}. For the sake
of simplicity, in this work we use a kink
approximation 
\begin{eqnarray}
\label{profiles}
H(z)&= & \frac{1}{2}\, v(T)\,\left(1-\tanh\left[\alpha\left(1-
\frac{2\, z}{L_\omega}\right)\right]\right)\nonumber\\  \nonumber\\
\end{eqnarray}
\begin{eqnarray}
\beta(z)= \beta-\frac{1}{2}\, \Delta\beta
\left(1+\tanh\left[\alpha\left(1-
\frac{2\, z}{L_\omega}\right)\right]\right)\nonumber\ .
\end{eqnarray}
This approximation has been checked
to reproduce the exact calculation of the Higgs profiles within a few percent
accuracy, 
provided that we borrow from the exact calculation the values of the
thickness $L_\omega/2\alpha$ and the variation of the angle 
$\beta(z)$ along the
bubble wall, $\Delta\beta$, as we will do. In particular we will take
$\alpha= 3/2$, $L_\omega=20/T$, and we have checked that 
the result varies
only very slowly with those parameters,
while we are taking the values of $\Delta\beta$ which are obtained
from the two-loop effective potential used in 
our calculation.
\begin{figure}[htb]
\vspace{9pt}
\includegraphics*[scale=0.28]{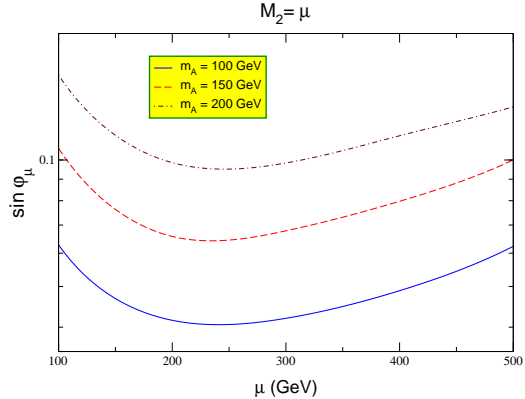}
\caption{Plot of $\sin\varphi_\mu$ as a function of $M_2=\mu$ for the
values of supersymmetric parameters specified in the text.}
\label{fig:plot2}
\end{figure}

In Fig.~\ref{fig:plot2} we have fixed $\eta=\eta_{\rm BBN}$ and plot 
$\sin\varphi_\mu$, where $\varphi_\mu$ is defined as 
$\mu_c=\mu\exp(i\varphi_\mu)$, as a function of $\mu$. We have fixed all
bubbles and supersymmetric parameters as in Fig.~\ref{fig:sources}, 
fixed $M_2=\mu$ and ran
over three typical values of the pseudoscalar Higgs mass $m_A$. In all cases
the phase transition is strong enough first order, $v(T_c)/T\simeq 1$,
the running mass of the lightest stop is around 120 GeV and the 
Higgs mass is, within the accuracy of our calculations, between 110 
and 115 GeV.

Since the computed $\eta$ behaves almost linearly in $\sin\varphi_\mu$, and
we have fixed $\eta=\eta_{\rm BBN}$, the more baryon asymmetry is generated the
smaller the value of $\sin\varphi_\mu$. This can be seen from 
Fig.~\ref{fig:plot2} where we have been working at the resonance peak
$M_2=\mu$ and hence baryogenesis has been maximized. We see, for all values 
of $m_A$, that the minimum of $\sin\varphi_\mu$ sits around $\mu\simeq 200$
GeV. The value of $\sin\varphi_\mu$ at the minimum decreases with $m_A$. 
In particular, for $m_A\simeq 100$ GeV, which is about the lower limit from
present LEP data (in fact, the present preliminary bounds from 
LEP~\cite{Higgs} are
$m_A> 89.9$ GeV for large values of $\tan\beta$, as those we are using in our
plots) we obtain that $\sin\varphi_\mu\simgt 0.04$. The dependence of
$\sin\varphi_\mu$ with respect to supersymmetric and bubble parameters,
as $M_2\neq \mu$, $m_A$ and $v_\omega$ has been thoroughly analyzed in
Ref.~\cite{recent} where the reader can find it as well as details of the
calculation of supersymmetric chargino sources. 

\section{Conclusions}

The main conclusion after a detailed analysis of both the phase transition and
the baryogenesis mechanism in the MSSM is that it is still alive after the
recent experimental results at high-energy colliders and, in particular, at 
the LEP collider.

Concerning the phase transition, its strenght is controlled mainly by the
Higgs mass and the lightest (right-handed) stop mass. Bubbles are formed
with thick walls (the value of the thickness is $\sim L_\omega/3$, with
$L_\omega\sim 20/T$) and propagate with extremely non-relativistic 
velocities ($v_\omega\sim 0.1-0.01$). The strenght of the phase transition has
to be such that $v(T)\simgt T$ at the critical temperature. This imposes a
strong constraint on the supersymmetric parameters in order to avoid
sphaleron erasure, in particular in view of the most recent (preliminary)
bounds on the SM-like Higgs mass, $m_H>113.2$ GeV and the observed
excess of events with $b\bar b$ invariant mass $\sim$ 114 GeV. Here two
possibilities can be drawn:
\begin{itemize}
\item
The observed excess of events corresponds to a Higgs signal.

\noindent
In this case the combined Higgs mass and BAU requirements impose some
restrictions on the supersymmetric parameters. {\bf a)} Heavy pseudoscalars
and large $\tan\beta$: say $m_A>150$ GeV and $\tan\beta>5$; {\bf b)} 
Heavy left-handed stops and controlled stop mixing: $m_Q\simgt 1$ TeV and
$0.25\simlt A_t/m_Q \simlt 0.4$; and, {\bf c)} Light right-handed stops:
$105\, {\rm GeV}\simlt m_{\widetilde t}\simlt 165\, {\rm GeV}$. In this case
the first prediction of the BAU scenario would have been realized and we would
need confirmation for the rest, in particular for the light stop.

\item
The observed excess of events does not correspond to a Higgs signal.

\noindent 
In that case a reduction of the $H b\bar b$ coupling would be needed.  
For the values of $A_t$ and $\mu$ consistent with electroweak baryogenesis,
a reduction of the coupling of the CP-even Higgs boson to
the bottom quark would demand not only small values
of $m_A \simeq 100$--150 GeV, but also large values
of $\tan\beta > 10$ and of $|\mu A_t|/m_Q^2 > 0.1$
(the larger $\tan\beta$, the easier 
suppressed values of the bottom quark coupling are obtained). A detailed
discussion on this issue has already been done~\cite{recent}.
\end{itemize}

Finally, 
concerning the generated baryon asymmetry, we have found that it requires
the CP-violating phase to be, $\varphi_\mu\simgt 0.04$. Values of 
$\varphi_{\mu} \simgt 0.04$ can  lead to acceptable phenomenology
if either peculiar cancellations in the squark and slepton 
contributions to the neutron and electron electric dipole moments (EDM) 
occur~\cite{cancelacion}, and/or if the first 
and second generation of squarks are heavy. This second
possibility is quite appealing and, as has been recently
demonstrated~\cite{DarkM}, 
leads to acceptable phenomenology.

\section*{Acknowledgments}
I would like to thank M.~Carena, J.M.~Moreno, M.~Seco and
C.E.M.~Wagner for the intensive collaboration on the subject during the last
year. This work has been supported
in part by CICYT, Spain, under contract AEN98-0816, and by EU 
under contract HPRN-CT-2000-00152.

\end{document}